\documentclass[prd,aps,showpacs,secnumarabic,superscriptaddress]{revtex4}
\usepackage[dvips]{graphicx,color}
\begin{document}
\def\ds{\displaystyle}
\def\ss{\scriptstyle}
\def\sb{\mbox{\rule{0pt}{8pt}}}
\def\hh{\mbox{\rule[-6pt]{0pt}{20pt}}}
\def\ti{\times}
\def\Om{\Omega}
\title{Comments on an alternative theory for the accelerating universe} 
\author{Duane A. Dicus}
\affiliation{Center for Particles and Fields, University of Texas, Austin, TX 78712} \email{dicus@physics.utexas.edu}
\author{Wayne W. Repko}
\affiliation{Department of Physics and Astronomy, Michigan State University, East Lansing, MI 48824} \email{repko@pa.msu.edu}
\date{\today}

\begin{abstract}
    We utilize the recently released supernova data of Riess {\it et al.} on the acceleration of the universe to determine how the parameters of a particular modification of gravity, the DGP 5-dimensional theory, are changed when compared to determinations using earlier data sets.  We show the parameters of the theory are now very tightly constrained and at very reasonable values.  For example, the curvature parameter, $\Omega_k$, in contrast to an earlier analysis which used a smaller data set, is now consistent with zero, and the redshift at which the universe starts to accelerate is near unity.
\end{abstract}
\maketitle

\section{Introduction}

Since the discovery that the expansion of universe is accelerating \cite{hiz,super}, most of the work has been in the direction of describing this acceleration by a negative pressure component  -- the dark energy -- within the usual theory of gravity. Alternative theories try to eliminate the dark energy by modifying the theory of gravity itself either by adding extra dimensions in which gravity, but only gravity, propagates  \cite{5d} or by modifying Einstein's equations in four dimensions\cite{Jose}. Here we will consider one such alternative theory, the so-called DGP model \cite{DGP}, where gravity is effectively 4-dimensional at short or moderate distances but 5-dimensional at large distances. An extensive review of the DGP theory has been given by Lue \cite{Lue}. The supernova data \cite{four} have been fit to this model with the result that the curvature was required to be very large \cite{Deff,AP}. As shown by Maartens and Majerotto \cite{MM}, once the comoving distance to the surface of last scattering given by WMAP \cite{WMAP,WMAP3} is included as a constraint on the data, the matter density and curvature agree with other measurements such as the latest results from WMAP \cite{WMAP3}. Here, we examine the implications of using the latest supernova data set of Riess {\it et al.} \cite{R06} to fit the parameters of the DGP model.

It is known that the form of the Hubble parameter in the DGP theory is indistinguishable from a dark energy model \cite{Deffonly,Deff}.  We check that the equation of state of this effective dark energy, $w(z)$, is well behaved -- it varies from $-0.8$ at small redshifts to $-0.5$ at large redshifts.  It has been argued, however, that the DGP theory allows values of the effective $w(z)$ that are less that $-1$ \cite{SS,LS};  we will discuss this point briefly. 

\section{Formalism and Results}

The DGP theory has only one parameter, $r_c$, the crossover radius where the theory changes between a region that is effectively 4-dimensional to one that is fully 5-dimensional, defined by
\begin{equation}\label{rc}
r_{c}\,=\,\frac{M_{Pl}^2}{2\,M_5^3}\,,
\end{equation}
where $M_5$ is the 5-d Planck mass. If we introduce a density parameter associated with $r_{c}$ as
\begin{equation}\label{Or}
\Omega_{r}\,=\,\frac{1}{4\,r^2_cH^2_0}\,,
\end{equation}
then the Hubble parameter, $H(z)$, is given by \cite{Deffonly,Deff}
\begin{equation}\label{H}
\frac{H^2(z)}{H^2_0}\,=\,\Omega_k(1+z)^2\,+\,\left[\epsilon\sqrt{\Omega_r}\,+ \,\sqrt{\Omega_r+\Omega_M(1+z)^3}\right]^2\,,
\end{equation}
where $z$ is the redshift and $\Omega_M$ and $\Omega_k$ are the matter and curvature density parameters. There are two possible phases denoted by $\epsilon=\pm 1$. For obvious reasons we will use the self-accelerating $\epsilon=+1$ phase, but we will have occasion below to refer to the other phase. From (\ref{H}) we must have
\begin{equation}\label{norm}
\Omega_k\,+\,\left[\sqrt{\Omega_r}\,+\,\sqrt{\Omega_r+\Omega_M}\right]^2\,=\,1\,.
\end{equation}
We will treat $\Omega_r$ and $\Omega_M$ as the independent parameters.

To confront this theory with the data we need the luminosity distances which are given by the usual integral of $1/H$. Prior to the recent supernova data release of Riess {\it et al.} \cite{R06}, the largest data sets available were the 164 supernova events collected in Ref.\cite{four} and Ref.\cite{essence}, the sum of which we call the essence data, and the legacy data set of Ref.\cite{legacy}. Fits of the data in Refs. \cite{four} and \cite{legacy} to DGP model were performed by Maartens and Majerotto \cite{MM}. We reproduce their results (including the data in Ref. \cite{essence}) for comparison with the fits to the data in Ref. \cite{R06}.

For the essence data, a $\chi^2$ fit varying $\Omega_r$, $\Omega_M$ and an offset gives
\begin{eqnarray}
\Omega_r\,&=&\,0.254^{+0.038}_{-0.044}\,,  \label{Or1}  \\
\Omega_M\,&=&\,0.348^{+0.073}_{-0.067}\,,  \label{OM1}
\end{eqnarray}
where the errors are one sigma. This agrees with the results of Ref.\cite{AP} which used the gold data \cite{four} and looks quite reasonable until we use Eq.(\ref{norm}) to find $\Omega_k$. The result is an unattractively large value
\begin{equation}\label{Ok1}
\Omega_k\,=\,-0.638^{+0.43}_{-0.36}\,.
\end{equation}
Alternately use of the legacy data of Ref.\cite{legacy} gives somewhat better numbers but with larger errors,
\begin{eqnarray}
\Omega_r\,&=&\,0.206^{+0.068}_{-0.082}\,, \label{Or2}  \\
\Omega_M\,&=&\,0.231^{+0.148}_{-0.173}\,, \label{OM2}  \\
\Omega_k\,&=&\,-0.244^{+0.93}_{-0.76}\,.        \label{Ok2}
\end{eqnarray}

\begin{figure}[h]
\begin{minipage}[t]{0.45\textwidth}
  \centering
  \includegraphics[width=3.0in]{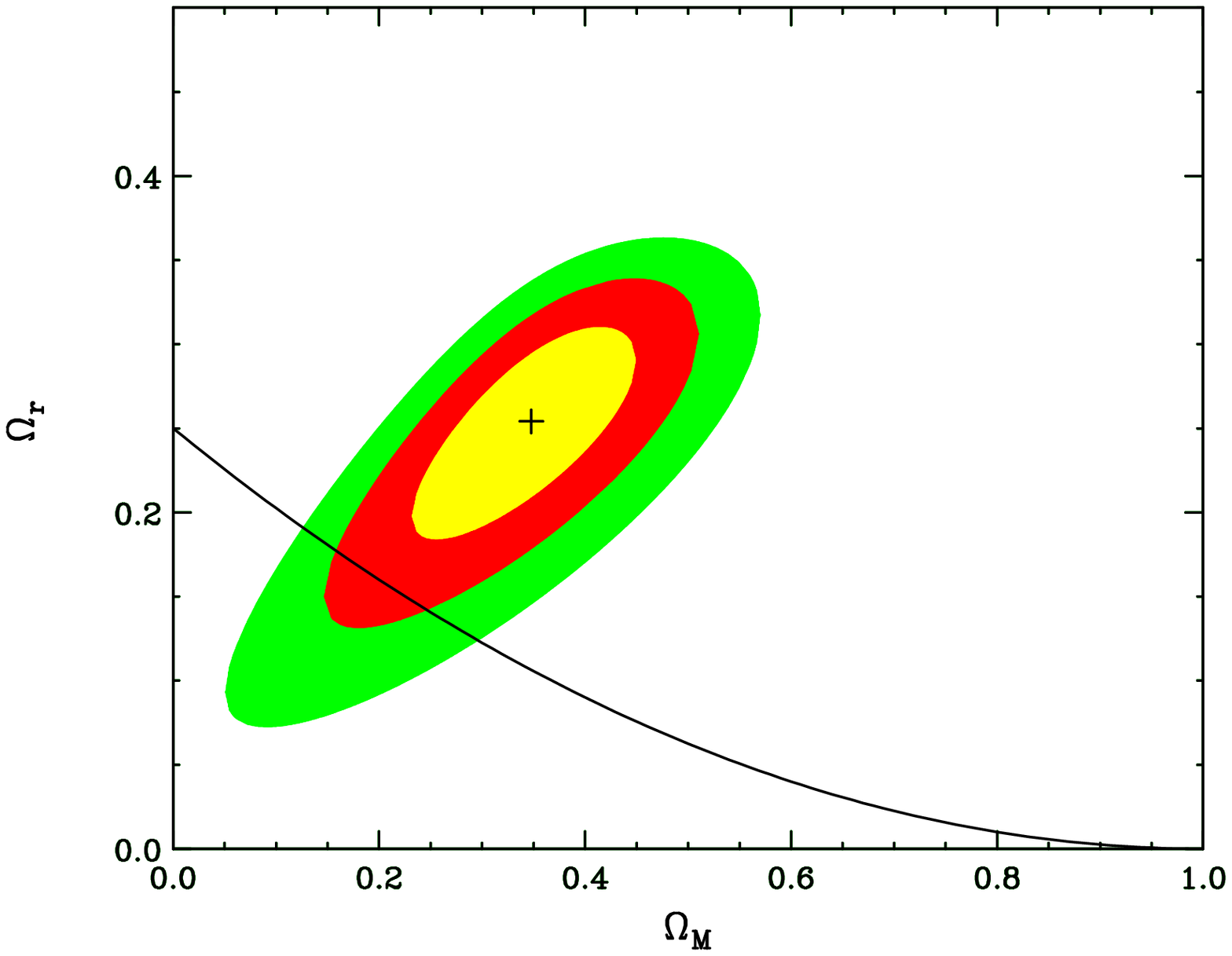}
  \caption{The $68.3\%$, $95.4\%$, and $99.7\%$ confidence contours in the $\Omega_M-\Omega_r$ plane are shown for the Hubble parameter of Eq.\,(\ref{H}) using the essence data without the CMB point. The cross denotes the $\chi^2$ minimum located at $\Omega_M=0.348$, $\Omega_r=0.254$, and the line is where $\Omega_k=0.0$.  \label{essbrane}}
\end{minipage}%
\begin{minipage}[t]{0.1\textwidth}
\hfil
\end{minipage}%
\begin{minipage}[t]{0.45\textwidth}
  \centering
  \includegraphics[width=3.0in]{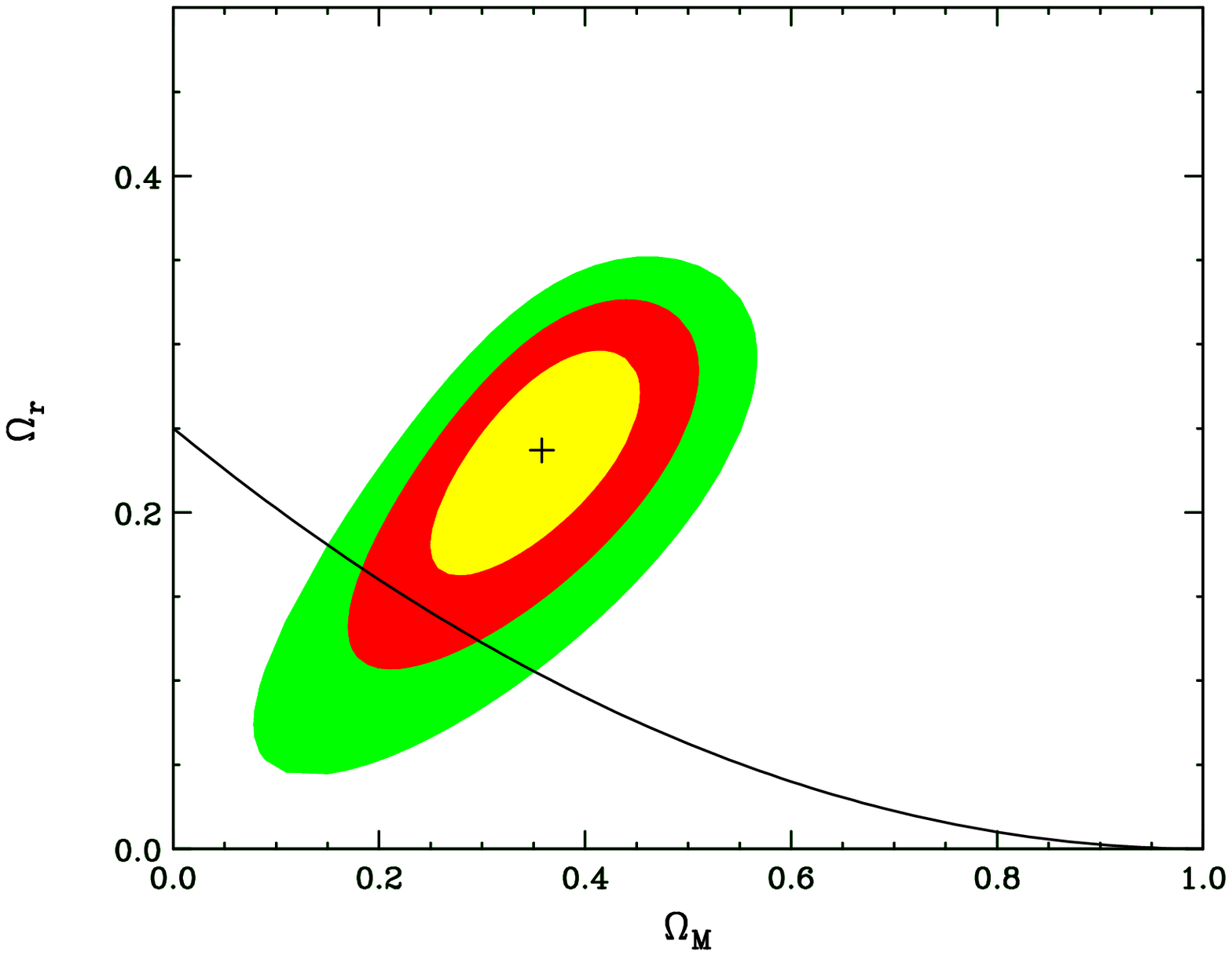}
  \caption{The $68.3\%$, $95.4\%$, and $99.7\%$ confidence contours in the $\Omega_M-\Omega_r$ plane are shown for the Hubble parameter of Eq.\,(\ref{H}) using the  R06 data with no priors. The cross denotes the $\chi^2$ minimum located at $\Omega_M=0.358$, $\Omega_r=0.237$, and the line is where $\Omega_k=0.0$. \label{R06brane}}
\end{minipage}
\end{figure}

However, there is one more luminosity constraint that we can add to the data sets and that is given by the comoving distance to the surface of last scattering measured by WMAP.  In Ref.\cite{WMAP} this is called the `angular size distance to the decoupling surface', $d_A$, and it is what Weinberg calls the proper-motion distance \cite{steve}.  In particular, we include this data point as a (Gaussian) prior on the parameter \cite{WT,BET} $R_{CBM}=\sqrt{\Omega_M}d_A = 1.70\pm 0.03$ at $z=1089$ \cite{WMAP}. Including this one additional point with its very long lever arm changes the results of Eqs.(\ref{Or1}), (\ref{OM1}), and (\ref{Ok1}) to
\begin{eqnarray}
\Omega_r\,&=&\,0.154^{+0.011}_{-0.012}\,,  \nonumber   \\
\Omega_M\,&=&\,0.191^{+0.034}_{-0.030}\,,  \label{OM3} \\
\Omega_k\,&=&\,0.039^{+0.032}_{-0.036}\,.  \nonumber
\end{eqnarray}
The other data set which gave Eqs.\,(\ref{Or2}), (\ref{OM2}), and (\ref{Ok2}) now gives
\begin{eqnarray}
\Omega_r\,&=&\,0.169^{+0.011}_{-0.011}\,,  \nonumber    \\
\Omega_M\,&=&\,0.150^{+0.030}_{-0.027}\,,  \label{OM4}  \\
\Omega_k\,&=&\,0.048^{+0.027}_{-0.031}\,.  \nonumber
\end{eqnarray}
These results are much better. The problem of the large negative $\Omega_k$ has gone away and the errors are much smaller.  The $\Omega_M$ value for the essence data looks a bit small but is consistent with the most recent WMAP \cite{WMAP3} results. In the case of the legacy data, the value of $\Omega_M$ is low. 

\begin{figure}[h]
\begin{minipage}[t]{0.45\textwidth}
  \centering
  \includegraphics[width=3.0in]{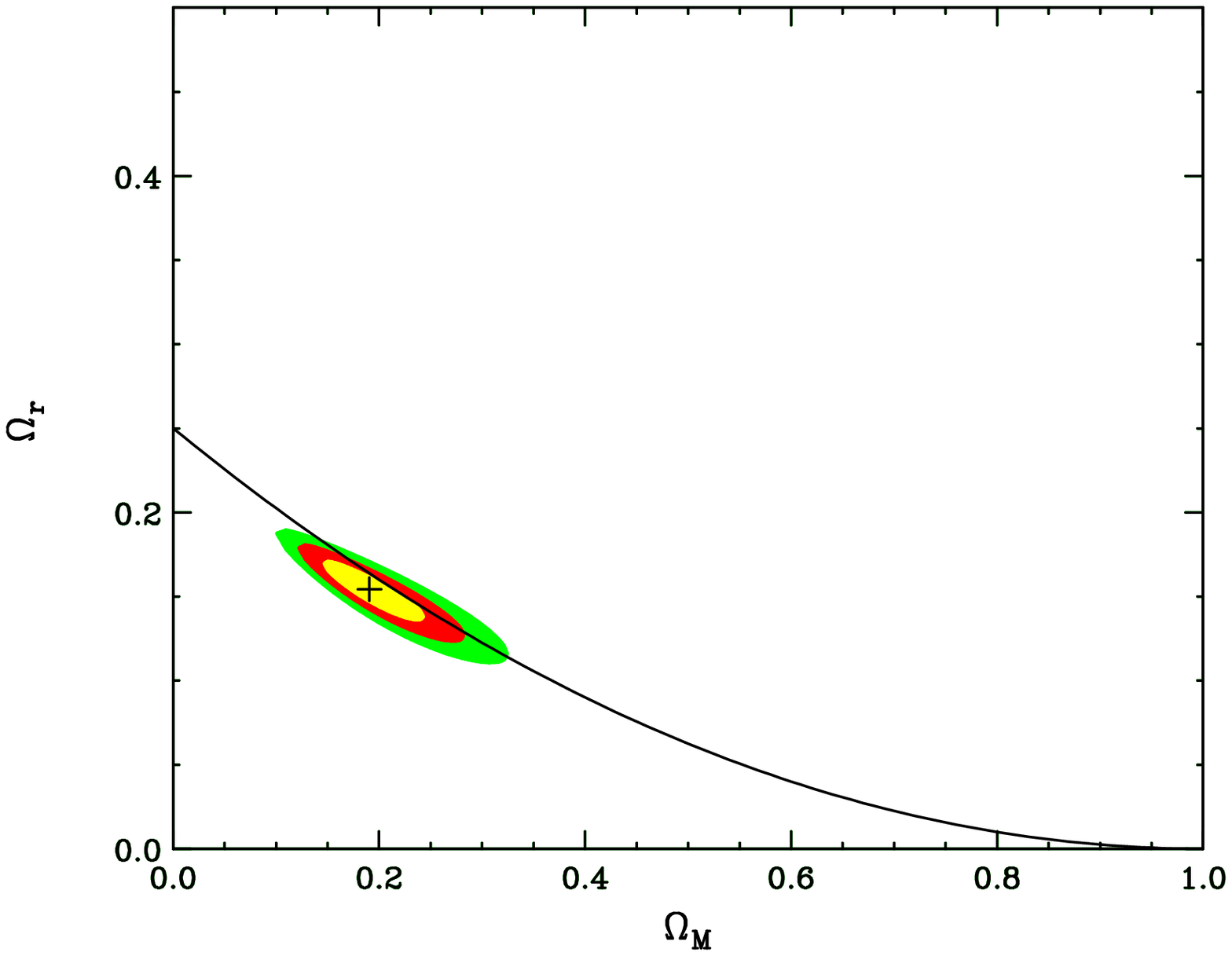}
  \caption{\footnotesize The $68.3\%$, $95.4\%$, and $99.7\%$ confidence contours in the $\Omega_M-\Omega_r$ plane are shown for the Hubble parameter of Eq.\,(\ref{H}) using the essence data and the CMB prior. The cross denotes the $\chi^2$ minimum located at $\Omega_M=0.191$, $\Omega_r=0.154$, and the line is where $\Omega_k=0.0$. \label{essbrane_CMB} }
\end{minipage}%
\begin{minipage}[t]{0.1\textwidth}
\hfil
\end{minipage}%
\begin{minipage}[t]{0.45\textwidth}
  \centering
  \includegraphics[width=3.0in]{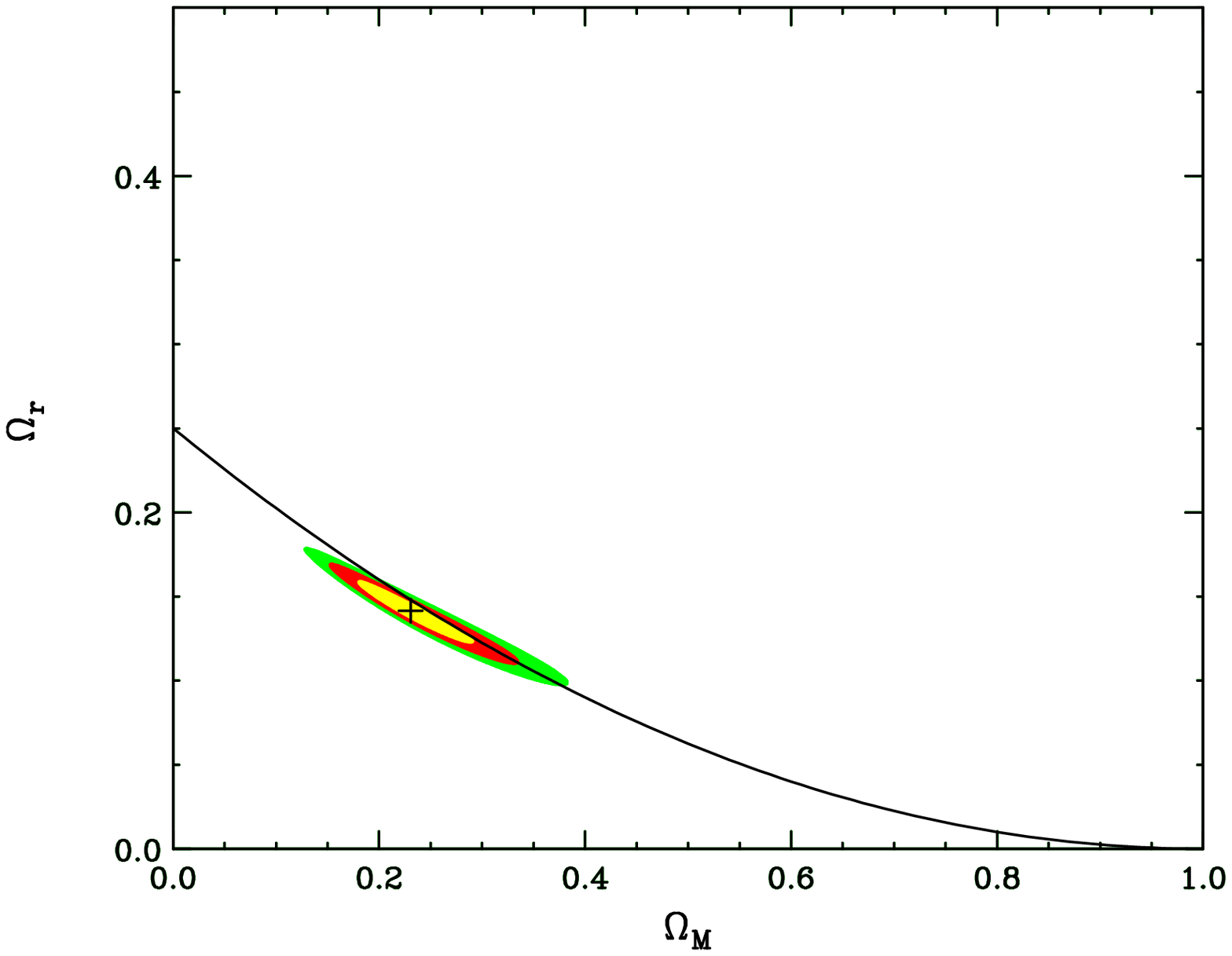}
  \caption{\footnotesize The $68.3\%$, $95.4\%$, and $99.7\%$ confidence contours in the $\Omega_M-\Omega_r$ plane are shown for the Hubble parameter of Eq.\,(\ref{H}) using the R06 data and the CMB prior. The cross denotes the $\chi^2$ minimum located at $\Omega_M=0.231$, $\Omega_r=0.142$, and the line is where $\Omega_k=0.0$.}\label{R06brane_CMB}
\end{minipage}
\end{figure}

Repetition of the these calculations with and without the CMB point using the 206 supernovae of Ref.\,\cite{R06}, which we will refer to as the R06 data, gives 
\begin{eqnarray}
\Omega_r\,&=&\,0.237^{+0.039}_{-0.046}\,, \nonumber   \\
\Omega_M\,&=&\,0.358^{+0.062}_{-0.067}\,, \label{OM5} \\
\Omega_k\,&=&\,-0.583^{+0.455}_{-0.357}\,,\nonumber 
\end{eqnarray}
and 
\begin{eqnarray}
\Omega_r\,&=&\,0.142^{+0.011}_{-0.012}\,,\nonumber   \\
\Omega_M\,&=&\,0.231^{+0.037}_{-0.037}\,,\label{OM6} \\
\Omega_k\,&=&\,0.027^{+0.020}_{-0.024}\,,\nonumber  
\end{eqnarray}
respectively. When performing these fits, we followed the procedure employed in Ref.\,\cite{R06} and did not include the 24 supernovae with $z<0.024$. Again, the R06 data without the CMB point favors a large value of $\Omega_M$ and a large negative value of $\Omega_k$. When the CMB point is included, $\Omega_M$ value is quite reasonable and $\Omega_k$ is small. The confidence contours for the essence data and the R06 data without the CMB point are compared in are Figs.\ref{essbrane} and \ref{R06brane}, and the corresponding comparison  with the CMB point included is shown in Figs.\,\ref{essbrane_CMB} and \ref{R06brane_CMB}. What is particularly striking is that in either case the inclusion of the CMB point tightly constrains the parameters $\Omega_M$ and $\Omega_r$ along the $\Omega_k=0$ line, with the R06 data preferring a wider range of values for $\Omega_M$. 

\begin{table}[h]
\centering
\begin{tabular}{| c | c | c | c | c | c |c|}
\hline
\hh $\Om_M$  & $\Om_r$  & $\Om_k$  & $\chi^2$  & MAT  & BAO  & CMB \\
\hline
\hh $0.358^{+0.062}_{-0.067}$ & $0.237^{+0.039}_{-0.046}$ & $-0.583^{+0.455}_{-0.357}$& $156.4$  &       &       &       \\
\hline
\hh $0.231^{+0.037}_{-0.037}$ & $0.142^{+0.011}_{-0.012}$ & $0.027^{+0.020}_{-0.024}$ & $160.2$  &       &       & $\ti$ \\
\hline 
\hh $0.281^{+0.018}_{-0.017}$ & $0.193^{+0.028}_{-0.030}$ & $-0.272^{+0.187}_{-0.166}$& $157.8$  & $\ti$ & $\ti$ &       \\
\hline
\hh $0.284^{+0.023}_{-0.022}$ & $0.125^{+0.008}_{-0.007}$ & $0.012^{+0.020}_{-0.022}$& $162.8$  &       & $\ti$ & $\ti$ \\
\hline
\end{tabular}
\caption{\footnotesize The results for fits to the R06 data to the DGP model are shown for various combinations of priors. MAT denotes the matter density prior $\Om_M=0.28\pm 0.03$, BAO denotes the baryon acoustic oscillation prior $A=0.469\pm 0.017$ and CMB denotes the WMAP prior $R_{CMB}=1.70\pm 0.03$. Note that for all cases $\chi^2/{\rm DOF}<0.91$.}\label{T1}
\end{table}

\begin{figure}[h]
\begin{minipage}[t]{0.45\textwidth}
  \centering
  \includegraphics[width=3.0in]{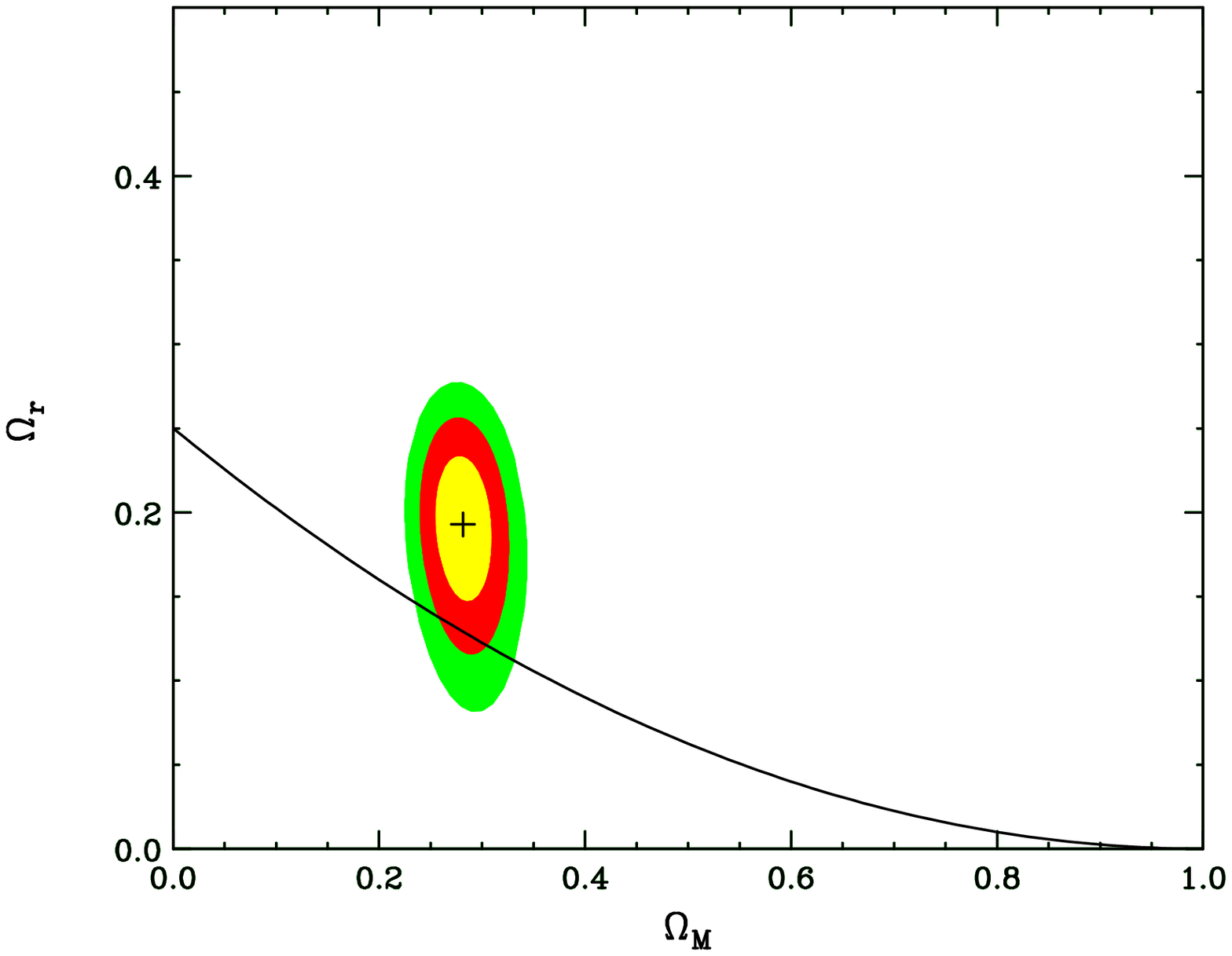}
  \caption{\footnotesize The $68.3\%$, $95.4\%$, and $99.7\%$ confidence contours in the $\Omega_M-\Omega_r$ plane are shown for the Hubble parameter of Eq.\,(\ref{H}) using the R06 data and the and the `weak' prior consisting of the the priors on $\Omega_M$ and the baryon acoustic oscillation parameter $A$. The cross denotes the $\chi^2$ minimum located at $\Omega_M=0.281$, $\Omega_r=0.193$, and the line is where $\Omega_k=0.0$. \label{weak} }
\end{minipage}%
\begin{minipage}[t]{0.1\textwidth}
\hfil
\end{minipage}%
\begin{minipage}[t]{0.45\textwidth}
  \centering
  \includegraphics[width=3.0in]{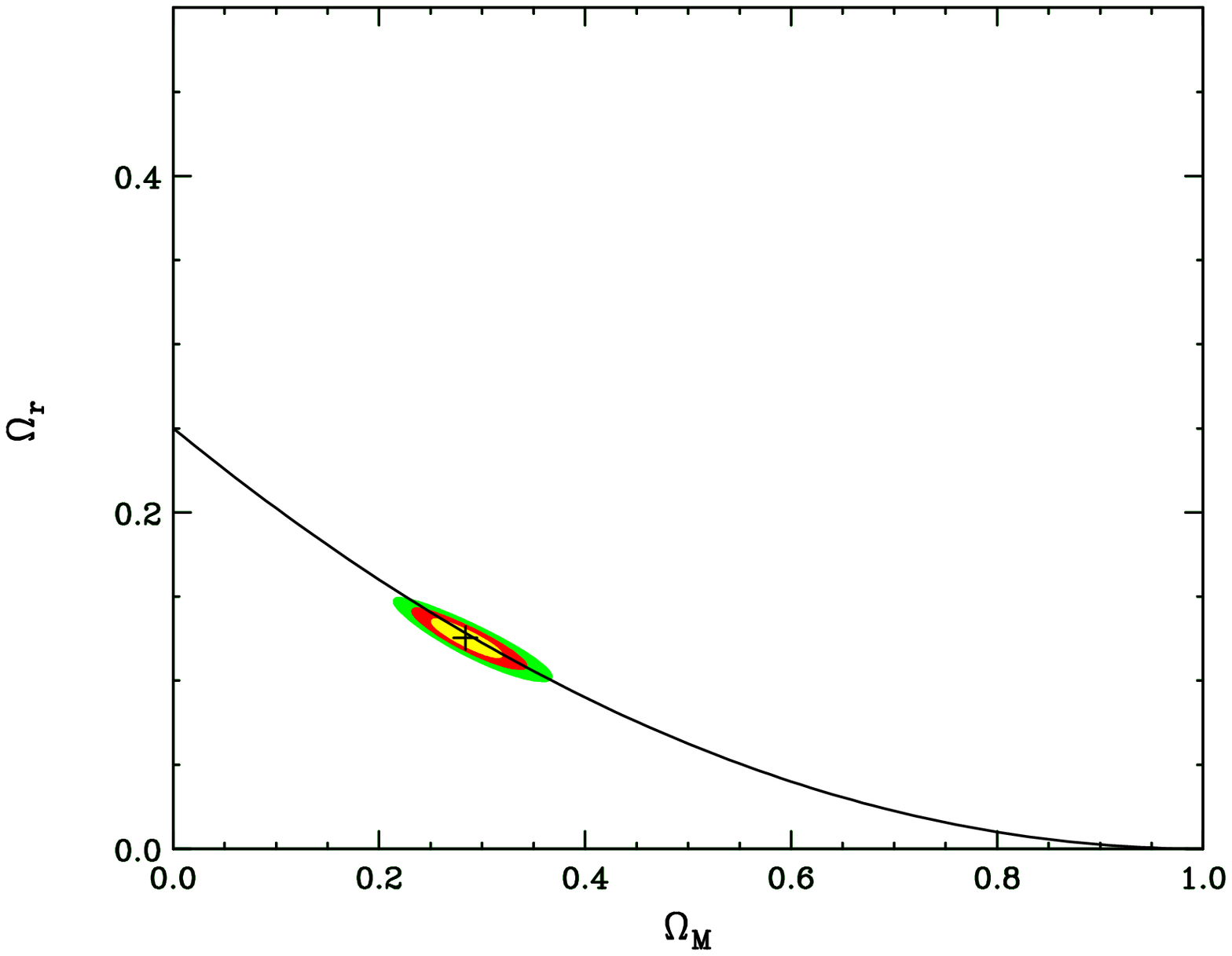}
  \caption{\footnotesize The $68.3\%$, $95.4\%$, and $99.7\%$ confidence contours in the $\Omega_M-\Omega_r$ plane are shown for the Hubble parameter of Eq.\,(\ref{H}) using the R06 data and the `strong' prior consisting of the priors on the baryon acoustic oscillation parameter $A$ and the CMB parameter $R_{CMB}$. The cross denotes the $\chi^2$ minimum located at $\Omega_M=0.284$, $\Omega_r=0.125$, and the line is where $\Omega_k=0.0$.}\label{strong}
\end{minipage}
\end{figure}

In Ref.\,\cite{MM}, fits to the DGP model using the gold data of Ref.\,\cite{four} included a prior on the baryon acoustic oscillation parameter (BAO) \cite{FG,E} $A=0.469\pm 0.017$ as well as the CMB constraint and, in the Ref.\,\cite{R06} analysis of the $\Lambda$CDM model, fits combining priors on $\Omega_M=0.28\pm 0.03$ and $A$ (weak prior) and $A$ and the CMB point (strong prior) were discussed. The implications of the weak and strong priors on the DGP model using the R06 data are tabulated in Table\,\ref{T1} and illustrated in Figs.\,\ref{weak} and \ref{strong}. In the case of the weak prior, the inclusion of a prior on mass parameter has the effect of orienting the confidence contours in such a way as to favor a narrow range for values for $\Omega_M$ as well as significantly reducing the size of the allowed parameter space. Interestingly, the strong prior, which only involves the baryon acoustic oscillation and CMB constraints, shifts the most probable value of $\Omega_M$ to about the same value as obtained with the weak prior, in addition to orienting the confidence contours in a narrow range along the $\Omega_k=0$ line. Including the extra 24 R06 events with $z<0.024$ changes the $\Omega_r$, $\Omega_M$, $\Omega_k$ values in Table\,\ref{T1} by at most 0.004 (and usually much less). The $\chi^2$ increases by about 13 in each case.

Given these very precise values for the parameters what do they predict for the redshift where the universe starts accelerating? The deceleration parameter can be written
\begin{equation}\label{q}
q(z)\,=\,-1+(1+z)\frac{H'(z)}{H(z)}
\end{equation}
where the prime indicates differentiation with respect to $z$. Using Eq.\,(\ref{H}) this gives, after some amount of work,
\begin{equation}\label{qND}
q(z)\,=\,\frac{N}{D}
\end{equation}
where, with $s=\Omega_r/\Omega_M$,
\begin{eqnarray}
D\,=\,\sqrt{s+(1+z)^3}\left\{\frac{\Omega_k}{\Omega_M}(1+z)^2+\left[\sqrt{s+(1+z)^3}+\sqrt{s}\right]^2\right\}  \label{D}  \\
N\,=\,\frac{1}{2}(1+z)^3\left[\sqrt{s+(1+z)^3}-\sqrt{s}\right]\,-\,2s\left[\sqrt{s+(1+z)^3}+\sqrt{s}\right]   \label{N}
\end{eqnarray}
Since $\Omega_k>0$ the denominator is always positive.  Then it is not too hard to see that the redshift where $N$ is zero, $z^{*}$, is given by
\begin{equation}\label{zstar}
1\,+\,z^{*}\,=\,2s^{\frac{1}{3}}
\end{equation}
For the essence parameters in Eqs.(\ref{OM3}) we find $z^{*}=0.862\pm 0.114$.  Using the legacy parameters of Eqs.\,(\ref{OM4}) the value is slightly higher, $z^{*}=1.082\pm 0.137$.  The result for the R06 parameters, Eqs.\,(\ref{OM6}), is $z^* = 0.701\pm 0.100$. Note that Eq.\,(\ref{zstar}) also tells us that $s$ must be larger than $\frac{1}{8}$ for acceleration to have started by today.

Now since the Hubble parameter of this theory is known to be identical with that of dark energy we want to see what form the dark energy equation of state takes.  We will compare the  dark energy with the general form of Eq.(\ref{H}) which allows for both phases.
For a dark energy description, the Hubble parameter is given by
\begin{equation}\label{HDE}
\frac{H^2(z)}{H^2_0}\,=\,\Omega_k(1+z)^2+\Omega_M(1+z)^3+\Omega_Xf(z)
\end{equation}
where $\Omega_Xf(z)$ is the dark energy density and $f(z)$ is given in terms of the dark energy equation of state by
\begin{equation}\label{f}
f(z)\,=\,\exp\left\{3\int_0^z\frac{dx}{1+x}(1+w(x))\right\}
\end{equation}
Comparing Eqs.(\ref{H}) and (\ref{HDE}) we see that
\begin{equation}\label{Of}
\Omega_Xf(z)\,=\,2\Omega_r\,\pm\,2\epsilon\sqrt{\Omega_r}\sqrt{\Omega_r+ \Omega_M(1+z)^3}
\end{equation}
From Eq.(\ref{f}) we see that $w(z)$ is given by
\begin{equation}\label{w}
w(z)\,=\,-1\,+\,\frac{1}{3}(1+z)\frac{f'(z)}{f(z)}
\end{equation}
where, as above, the prime indicates differentiation with respect to $z$.  Applying this equation to Eq.(\ref{Of}) we get \cite{Deff}
\begin{equation}\label{w1}
w(z)\,=\,-1\,+\,\frac{1}{2}\frac{(1+z)^3}{s+(1+z)^3+\epsilon\sqrt{s}\sqrt{s+(1+z)^3}}
\end{equation}
where $s$ was defined above.  For both phases the second term on the righthand side of Eq.\,(\ref{w1}) is positive for all $s>0$ so $w(z)$ is always greater than $-1\,$. With $\epsilon=+1$ both of the above fits give $w(z)\approx -0.85$ at $z=0$ and  $w(z)\to -0.5$ at large $z$.

It has been claimed that the phase with the minus sign in (\ref{H}) can give $w(z)$ less than $-1$ \cite{SS,LS}. Rather than Eq.(\ref{H}) the authors of Ref.\cite{LS} use
\begin{equation}\label{HLS}
\frac{H(z)}{H_0}\,=\,-\sqrt{\Omega_r}+\sqrt{(1+\sqrt{\Omega_r})^2+ \Omega_M(1+z)^3-\Omega_M}
\end{equation}
Note that the righthand side of Eq.\,(\ref{HLS}) is unity when $z = 1$ so $\Omega_r$ and $\Omega_M$ can be treated as  independent variables. Now we can compare Eq.\,(\ref{HLS}) with Eq.(\ref{HDE}) with $\Omega_k$ set equal to zero, determine $\Omega_X\,f(z)$, and use Eq.(\ref{w}) to find $w(z)$.  The general expression is not very pretty but at $z =0$ it reduces to
\begin{equation}\label{w0}
w(0)\,=\,-1-\frac{\sqrt{\Omega_r}\,\Omega_M}{(1+\sqrt{\Omega_r})(1-\Omega_M)}
\end{equation}
which is less than $-1$ for $\Omega_M < 1$. We  tried to fit the data to this form of the Hubble parameter although we didn't expect much success since Eq.\,(\ref{HLS}) does not have the self-accelerating phase.  We left out the WMAP constraint because Eq.\,(\ref{HLS}) is only intended to work for small $z$. The data tried to force $\Omega_r$ and $\Omega_M$ to be large, $\Omega_r > 10$ and $\Omega_M > 1$, in which case $w(0)$ is not less than $-1$.  When we added a prior of $\Omega_M=0.30\pm0.03$ then $\Omega_r$ is very small, consistent with zero, $0.015\pm0.023$, but $w(0)$ is indeed less than $-1$.  

\section{Conclusions}

We have considered the DGP theory of gravity and shown that if the expression for the Hubble parameter given by Eq.\,(\ref{H}) holds for all redshifts back to the surface of last scattering so that the comoving distance measured by WMAP can be added as a constraint on the supernova data, then the parameters $\Omega_r,\, \Omega_M$ and $\Omega_k$ can be precisely predicted. Our results are given in Eqs.\,(\ref{OM3}) for the essence data, Eqs.\,(\ref{OM4}) for the legacy data, and in Table\,\ref{T1} for the R06 data. The effect of adding the CMB point is illustrated by the confidence contours for the essence and R06 data shown in Figs. \ref{essbrane}\,-\,\ref{R06brane_CMB} and the effect of using the weak and strong priors with the R06 data is shown in Figs.\,\ref{weak} and \ref{strong}.

This theory is consistent with a flat universe which began accelerating at a redshift equal to, or slightly less than, unity as given by Eq.\,(\ref{zstar}) and is identical, as far as the deliberations of this paper are concerned, with a model of dark energy whose equation of state is given by Eq.\,(\ref{w1}) with $s\,\approx\,0.6\,-\,1.1$.

\begin{acknowledgments}
We wish to thank Professor R. Maartens informing us of his work with E. Majerotto on this same subject \cite{MM}. This work was supported in part by the U.~S. Department of Energy under Grant No. DE-FG03-93ER40757 and by the National Science Foundation under Grant PHY-0244789 and PHY-0555544.
\end{acknowledgments}

\end{document}